\documentstyle[prl,aps,twocolumn]{revtex}
\def\break#1{\pagebreak \vspace*{#1}}
\begin{document}

\draft
\pagestyle{empty}

\title{ PEDESTRIAN NOTES ON QUANTUM MECHANICS}
\author{Haret C. Rosu 
}
\address{ \begin{center}
Instituto de F\'{\i}sica de la Universidad de Guanajuato, Apdo Postal
E-143, Le\'on, Gto, M\'exico\\
Institute of Gravitation and Space Sciences, Bucharest, Romania\\
rosu@ifug3.ugto.mx\\
(received: March 22, 1997;
published by Metaphysical Review 3, 8-22; May 1, 1997)
\end{center}
}

\maketitle
\widetext
{\scriptsize
`` Get your facts straight, and then you can distort them as much
as you please.''

Mark Twain}

\vskip 0.5cm

\begin{abstract}

  I present an elementary essay on some issues related to foundations
  of nonrelativistic quantum mechanics, which is written in the spirit
  of extreme simplicity, making it an easy-to-read paper. Moreover, one
  can find a useful collection of ideas and opinions expressed by many
  well-known authors in this vast research field.

\end{abstract}

\narrowtext

\section{Indefinables}

Physics is first of all the science of measurement. As Lord Kelvin
put it

\begin{quote}

I often say that when you can measure what you are speaking
about, and express it in numbers, you know something about it.

\end{quote}

According to Kelvin a collection of thoughts cannot advance to the
stage of Science without numbers. Any observable of interest in physics
should be measurable or expressed in terms of measurable quantities.
Length and time are two of the indefinables of classical mechanics,
 since on an intuitive base there
are no simpler or more fundamental quantities in terms of which
length and time may be expressed. The problem of space-time picture
of the physical world is connected with the rigour of exact description
of nature requiring, say, differential equations, by means of which we are
 able to gauge the intuitive space-time scales of any motion.
The full number of indefinables in
mechanics is three, as all its quantities could be expressed by only
three indefinables. The third mechanical indefinable is usually the
\underline{mass}, but also \underline{force} may be chosen \cite{dim}.
 Human beings in their everyday
lives are continuously ``measuring" the mechanical indefinables,
 as well as other indefinables of physics, e.g.,
the temperature, by means of their physiological senses.
 Of course, it is a very rough ``measurement", because
it could be expressed in words, not in numbers.
 Words and numbers are complementary units of knowledge. Pure numbers
 do not tell us much about Nature unless we assign them some
significance. As a good example consider the number 3.52. Just a (real)
number as any other. But now write it as $2\pi/e^{\gamma}$. \break{1.3207in}
For some physicists it has already a meaning.
Finally, let us
write down the full chain, i.e.,
$2\Delta_{0}/T_{c}=2\pi/e^{\gamma} =3.52 $.
It tells us that 3.52 is the BCS value for the ratio between the gap
at zero temperature and the critical temperature for the transition to
the superconductor phase. One gets 3.52 only in BCS theory.

Because of its beauty, I am temted to give a second example which has
been quoted by Noyes \cite{001} from the books of
 Stillman Drake on Galileo.
So, what about the number 1.1107...
Nothing special at first glimpse. But now let us give a first
significate: $1.1107...=\pi/2\sqrt{2} $.
Geometrically it is the ratio between the quarterperimeter of the
circle and the side of the square inscribed into that circle.
Geometrical (i.e., spatial) measurements and thinking
 were much developed by Old
Greeks. But Galileo was the first to give 1.1107... as the ratio
of two times, namely the time $t_{p}$ it takes for a pendulum of a
specific length $l$ to swing to the vertical through a small arc to
the time $t_{f}$ it takes a body to fall from rest through a distance
equal to the length of the pendulum.
Galileo's measurement was 942/850 = 1.108, but he was not aware he
measured $\pi/2\sqrt{2}$. However he gave a remarkable formulation
of the ``law of gravity". The Galilean gravitation states that the
ratio of the pendulum time to the falling time as specified above
is the constant 1.108, ``anywhere that bodies fall and pendulums
 oscillate".

 To obtain the number of
indefinables (NOI) a community of physicists should adopt rules of their
measurements, especially since NOI is not a fixed number, and new types
of experiments might augment NOI.
The rule of classical indefinables is to choose a durable standard
of unit for each of them and to have a good dividing engine. This has
been achieved rather easy for the meter of length but not so easy in
the case of the meter of time (second). In the latter case the great
difficulty has been for a long time the missing of an accurate dividing
machine. Large errors were continuously accumulating over historical epochs,
and people in the fields of Religion and Politics were forced
 to apply corrections at some times.
Atomic standards (lasers) have been introduced since early sixties
having a natural atomic time- dividing machine ( the atomic frequency).
However, even the very precise atomic standards display statistical
results \cite{a1}, and there are also reported abnormalities during
solar eclipses \cite{a2}.
At present,
interferometers could be used for dividing purposes too, and computing
machines are usually attached to measuring devices for a more rapid
conversion of the physical interactions into real numbers. As regarding
computers, there should be a continuous effort to study their rate of
producing numbers which is not depending only on the used algorithm, and
to have more involved definitions of computable numbers (so-called
Turing Problem \cite{t}).

Establishing measurement rules for indefinables is
extremely important for the conceptual constructions in the realm
 of physics. It is not at all an easy job in the case of quantum
 indefinables.
Since the microscopic world is described by another kind of mechanics,
the celebrated quantum mechanics, it may be that new indefinables come into
play. One new indefinable is the wavefunction, also called
the state vector \cite{v}, or, more realistically, the wavepacket.
These are concepts
essentially of mathematical origin, and hence {\em a priori}
 calculable ones. In a certain sense they correspond to the
fundamental indefinable of euclidean geometry- the point. The
geometrical point has neither dimensions nor any attached units,
but we can
assign numbers (coordinates) to it. In this way one may come to the
conclusion that quantum mechanics is more a mathematical theory
rather than a physical one. It is a ``wave" mechanics allowing
a corpuscular duality. The measurement problem will be {\em ab inizio}
extremely delicate for quantum mechanics, just because it is a
theory containing unmeasurable (mathematical) indefinables. To the
mathematical indefinables one could always assign generalized
probabilistic meanings depending substantially on the measuring scheme.
The mathematical and psycho/philosophical literature is extremely rich
in various axiomatic schemes in probability theory and its mental
implications \cite{prob}. Let us mention the so-called {\em belief functions}
on which Halpern and Fagin have recently elaborated
\cite{hf}.
These are functions that allocate a number between 0 and 1 to every subset
of a given set of objects. Such functions have been introduced by Dempster
in 1967 \cite{de}, and it would be quite interesting to have a quantum
mechanics based on them, e.g., to write down a {\ em belief} density matrix.

It became common lore to say
that the measurement process is a more or less instantaneous effect
inducing a reduction/collapse of the wavepacket, which is interrupting its
quantum unitary evolution as described by the Schr\"{o}dinger
equation. As far as the meaning of measurement is not quite clear
whenever one is dealing with probabilistic concepts, the whole
quantum theory is subject to severe questions of interpretation and
therefore is open to deep philosophical problems. The
axioms of the standard probability theory are not fixed for ever.
It is a fundamental scientific goal to exploit various modifications
of the probabilistic axioms rendering possible new interpretations
of quantum mechanics \cite{kh1}.

\section{The concept of massiveness}

Let us start this section with an excerpt from Glimm \cite{gl}
\begin{quote}

Between quantum length scales (atomic diameters of about $10^{-10}$ m)
and the earth's diameter ($10^{6}$ m) there are about 16 length scales.
Most of technology and much of science occurs in this range. Between the
Planck length and the diameter of the universe there are 70 length scales.
70, 16, or even 2 is a very large number. Most theories become
intractable when they require coupling between even two adjacent length
scales. Computational resources are generally not sufficient to resolve
multiple length scales in 3 dimensional problems and even in many
two-dimensional problems.

\end{quote}

At the present time, {\em technology} is penetrating into the nanometer
 scale \cite{01}, and even atomic scale \cite{a}.
  By {\em technology} one should understand: i) machine
 tools (i.e., processing equipment), ii) measuring instruments
 (inspection equipment), iii) super-inspection factors (e.g.,
well-qualified human beings). We have already machine tools at the
nanometer scale \cite{03}, and one can process shapes down there at only one
order of magnitude away from the atomic scale.

Already at the end of 1959
Feynman \cite{rf} delivered a remarkable talk on manipulating
and controlling things on a small scale. As a matter of fact, human
beings are closer to atomic scales than, say, galactic scales, and besides,
to fabricate small things is an absolute technological requirement.
A list of reasons why we want to make things small was provided by
Pease \cite{pi} of which we cite: it is {\em fun}, smaller devices work
{\em faster},
smaller devices consume {\em less power}, smallness is intrinsically
{\em good},
it is scientifically {\em important}.
One can add to the list of Pease {\it small is beautiful} introduced by Fubini
and Molinari \cite{fm}.

Nonetheless, even within mesoscopic world
 the measurement problem will preserve its features.
We shall continue to establish correlations between a ``system"
observable and an ``apparatus" observable, i.e., we shall do our
measurements in the same common manner
\cite{1}. The apparatus should be massive with respect to the
system (massive not necessarily meaning of macroscopic size),
 and should have a ``pointer", which cannot be but in localized
 quantum
states. According to DeWitt, {\em massiveness} of the pointer
compared to the measured system is absolutely necessary to get
experimental results/outcomes. However, the concept of {\em massiveness}
is not elaborated by DeWitt (unless to say that $M_{ap}\gg m_{s}$).
At the nanometer scale, mesoscopic tips are by now in much
use making possible the measurement of Van der Waals forces between
the tip and the surface under investigation at distances smaller than
100 nm. That means forces in the range $10^{-6}-10^{-7}$ N are measured
either when the tip is moved or
 the surface is slowly approaching the tip \cite{tip}.

 Massiveness is related to localization, and may
be very helpful in distinguishing amongst various theories of quantum
evolution. The fundamental goal of this family of theories is to explain
in a unifying
way microscopic (quantum) objects and macroscopic ones. Since it is
reasonable to think that massiveness is indeed related to the localization
features of the system, it would be interesting to study in detail the
conditions under which various systems make the transition to massiveness.
By this transition which, to this day, is one of the most unclear
in quantum mechanics \cite{massv}, one should not mean in a compulsory
manner the various semiclassical ($\hbar\rightarrow 0$; or better
$\hbar /m \rightarrow 0$; notice K.R.W. Jones \cite{krw} who showed
that one can also keep $\hbar$ fixed) approximations to
quantum mechanics or the transition to macrophysics $N\rightarrow\infty$.
  It is merely a transition in
 the sense of the Born-Oppenheimer approximation. This approximation,
 going back to 1927, refers to molecular wave functions and is essential
 in the interpretation of molecular spectra. It is a perturbation
 expansion in a small parameter defined as the fourth root of the
 electron mass divided by the mass of the nuclei,
$\kappa=(m/M)^{1/4}$. Denote by $H$ a common Hamiltonian for a system
of heavy and light particles and by $H_{cM}$ the hamiltonian of the
center of mass motion. The problem is to find out the manner in which
the eigenvalues W$_{i}$, and
eigenfunctions $\psi $ of $H^{'}=H-H_{cM}$ depend on $\kappa$
   \cite{rs}. The main hypothesis of Born- Oppenheimer is the existence
   of an equilibrium position $X_{0}$ of the heavy particles such that
 the eigenvalues W$_{i}$ and the scaled eigenfunctions,
 $ \phi (\xi ,x) =x^{3/2(N-1)} \psi (x\xi +X_{0}, x) $,
 are analytic in $\kappa ^{2}$ and $\kappa $, respectively, for fixed $m$
 in the neighbourhood of zero.

Very useful would be to consider the transitions to macroscopic world and
to massiveness as
problems with multiple time scales, which are pervading many
areas in applied mathematics \cite{mts}.

Let us end this section with the relationship between {\it big} and
{\it small} in quantum mechanics. For this, I shall present excerpts
from Dirac's {\em Principles} \cite{dir}.
In the first chapter, ``The Principle of Superposition",
Dirac states

\begin{quote}

So long as {\it big} and {\it small} are merely relative concepts,
it is no help to explain the big in terms of the small. It is
therefore necessary to modify classical ideas in such a way as to
give an absolute meaning to size.

\end{quote}

On the same page one can read

\begin{quote}

We may define an object to be big when the disturbance accompanying
our observation of it may be neglected, and small when the disturbance
cannot be neglected. This definition is in close agreement with the
 common meanings of big and small....In order to give an absolute
 meaning to size, such as is required for any theory of the ultimate
 structure of matter, we have to assume that {\it there is a limit
 to the fineness of our powers of observation and the smallness of
 accompanying disturbance - a limit which is inherent in the nature
 of things and can never be surpassed by improved technique or
 increased skill on the part of the observer}.

\end{quote}

\section{Quantum mechanics and diffusions}

Words like \underline{particle} and \underline{motion} could be considered
also in the
class of undefined (primary) concepts \cite{2}. Electrons, neutrons,
neutrinos and other \underline{'ons} and \underline{'inos} could be only
names that we
accept because of their intuitive power. All the so-called elementary
particles which have been introduced in the last one hundred years can
be considered as particular forms of propagating (transport)
processes and energy carriers \cite{fhs}.
 Indeed, Schr\"{o}dinger equation is the basic equation of
quantum world, but diffusion equations are, no doubt, more general
 equations. We say `no doubt' just because already in 1933, Fuerth
 \cite{f}
 showed that Schr\"{o}dinger equation could be written as a diffusion
equation with an imaginary diffusion coefficient, $D_{qm}=i\hbar/2m$.
This imaginary diffusivity is vexing and many stopped the analogy
at this point. On the other hand, negative diffusivity is more natural
and one may encounter it in multicomponent systems, implying local
increase in the energy of the system as discussed by Ghez \cite{3}.
Let us consider
a one-dimensional Schr\"{o}dinger equation
$ i\hbar\frac{\partial \psi}{\partial t} =H\psi $
where $ H= \frac{p^{2}}{2m}+ V  $ and
$p=-i\hbar \frac{\partial}{\partial x} $.
The diffusive character of such an equation for $\psi$ is obvious if
we take into account a source term related to the potential energy,
and the momentum playing the role of flux. For historical reasons the
diffusion interpretations ( they may come in three classes: in configuration
space, in phase space, and in imaginary time) were not favored during a
long lapse of time, though
today mainly because of the impetus provided by {\em quantum optics}, we
became accustomed with such methods as quantum-jump \cite{molm} and
quantum-state diffusion \cite{gper} to simulate dissipation processes.
Indeed, Schr\"{o}dinger obtained his
wave mechanics by means of a more intuitive analogy in which he put together
the Hamilton-Jacobi theory, relating geometrical optics and particle dynamics,
with de Broglie's matter waves. One could say that what
Schr\"{o}dinger did was to randomize a purely classical theory by
means of de Broglie hypothesis. It was a way of randomizing within the
classical formalism, but, more generally, one should be aware of the
multitude of randomization procedures \cite{sc}.

 The apparent difficulty
 of imaginary diffusivity is not essential when interpreting it
 in the proper way. The result can well be a more general theory.
 The picture of the {\it World} is that of an infinite number of clusters
 in the sense of percolation theory. Classes of clusters could be defined in
 terms of their relative diffusivities and fluxes. Some of them are ``static"
 relative to other more kinetic ones \cite{s}. In this diffuson context,
 the imaginary
 character of the diffusion coefficient for quantum particles is related to
 the passage from a parabolic differential equation to a
 hyperbolic one \cite{nagy}.

Even a presentation of the one dimensional diffusion equation, first in the
discretized form and then in the continuous limit, on the lines of the
{\em Primer} of Ghez \cite{3} is very helpful to understand the diffusive
aspects of the Schr\"odinger equation, and I recommend the reader to look in
that book. The toy model of Ghez
is a pedagogical isotropic one-dimensional random work, in which one consider
points on a line with an arbitrary fixed origin.
 For the passage to the continuum limit one must introduce a jump
 distance between the points and a continuous particle distribution,
 depending not only on time but also on the space variable such that
 to coincide at the discrete sites with the discrete particle distribution.

There are also papers dealing with the connection
between a classical Markov process of diffusion type and the quantum
mechanical form of the Hamiltonian for a classical charged particle in an
electromagnetic field \cite{rec}.
These two problems are equivalent as far as one is concerned with
the expectation values for the particle energies in the two cases.
Consider a continuity equation of the type
$\frac{\partial \rho}{\partial t} = -\nabla\cdot (\rho v) $
where
$v=v_{0}- D\nabla \ln \rho $.
Such a  continuity equation is in fact a Fokker-Planck
 equation for the probability density $\rho $ for the position  vector
 of a particle following a Markov process of diffusion type with
 diffusion coefficient $D$.
The expectation value for the energy of the particle
$ <E> = \int [mv_{0} ^{2} /2 +eV]\rho d^{3} x$
can also be written
$ <E>=
 \int [mv^{2} /2 + D^{2}m/2 (\nabla \ln \rho )^{2} +eV]\rho d^{3}x$, and
the connection with the electromagnetic phenomena can be established by
means of the celebrated Helmholtz theorem for a vector (considering the
velocity as a vector, thus no type of spin)
 $ v=\alpha \nabla \phi + \beta A $,
 where $\phi $ and $A$ are defined in the usual way up to a gauge
 transformation, and $\alpha $ and $\beta $ are constants which
 should be chosen in an appropriate way to achieve the correspondence.
There would be interesting to study the passage ways from microscopic
to macroscopic
description of electromagnetic fields \cite{im} in this framework.
The traditional one, going
back to Lorentz, and which is applicable to common molecular media,
is by averaging the differential equations for microscopic quantities
by integration over some macroscopic volume. This is the most trivial
procedure for going to the macroscopic approximation. There are other
approaches, e.g., the topological one as discussed by Brusin \cite{br}.

\section{Quantum mechanics and localizations}

The collapse postulate of quantum mechanics is one of the most debatable
points in the conceptual base of this theory, being at the
same time the main desideratum for a modified quantum dynamics
\cite{as}. The collapse of
the state vector is required by the formal quantum theory of
 measurement. One must assure somehow the decoherence of the macroscopic
states of the apparatus in order to have a definite outcome for any
experiment involving quantum particles.
We still do not know if this decoherence is dependent on the
particular interaction and hence on the particular type of
measurement or is a universal feature of the transition from
microscopic to macroscopic behaviour. The first hypothesis is
called environmental (Zeh-Joos) localization \cite{4}.
On the other hand, the universal localization, also known
as spontaneous (or GRW) localization is due to Ghirardi, Rimini
and Weber \cite{g1}. It is difficult to decide between the two models.
In our opinion, they are not completely opposite ideas.
The dynamical (environmental) localization may be specific
  to the particular experiment, while the spontaneous localization might
be thought of as related to the transition to massiveness, which one would
  like to see as universal. In this way having different purposes,
the two standpoints are not contradictory. At the same time GRW
localization could be considered only as a special type of
 environmental localization at the scale given by the parameters of
the model. The point is that these parameters have been raised to the
level of fundamental constants of Nature by the authors.
Anyway, one must spell out explicit conditions allowing to
pass from a regime of continuous spontaneous (or dynamic) localization to a
discontinuous regime characteristic to the GRW localization. We recall that
in the GRW approach the N-particle wave function of the non-relativistic
Schr\"{o}dinger quantum mechanics (NRQM) is coupled to a normalized
Gaussian jump factor $ J_{GRW}(x)= K\exp (-x^{2}/2a^{2}) $.
  The frequency of the jumps and the localization constant are
considered as two new fundamental constants of Nature of the
following orders of magnitude $ \nu _{GRW} \sim 10^{-15} s^{-1} =
10^{-8}$ year$^{-1}$ and $ a\sim 10^{-5}$ cm.

The spontaneous localization implied by the GRW model might be tested
experimentally by means of mesoscopic phenomena, e.g., by looking for
instabilities of the mesoscopic growing
(thread-like, filamentary) patterns.
Recently, Kasumov, Kislov,
and Khodos \cite{kkk} observed the displacements of the free
ends of threads of amorphous hydrocarbons of 200-500 $\AA$ in width
and 0.2-2.0 $\mu$m in length relative to a fixed reference point on the
screen of a transmission electron microscope. The minimal displacements
were of about 5 $\AA$, and the observations were made in a regime of
stationarity of the threads, i.e., very low density of the beam current
($\sim$ 0.1 pA/cm$^{2}$). KKK observed random jumps of the free ends of the
carbon threads of 10-30 $\AA$ in length and of frequency of $\sim 1$ Hz.
They discussed possible reasons to induce vibrations, and came to the
conclusion that no classical external forces could explain the jumps.
They attributed them to the ``quantum potential", and to localizations of
GRW type, but the range of the observed parameters do not correspond to that
of the GRW ones.

What I would say is that the jumps or mesoscopic fluctuations of
the carbon threads are a kind of mesoscopic Brownian motion which damps
in time, being different from the microscopic quantum fluctuations
which never damps out.


{\em Moreover, if one takes into account the recent work of
Sumpter and Noid \cite{sn} the KKK results can be classified as
{\em red herrings}. Sumpter and Noid assigned the onset of positional
instabilities in samples of carbon nanotubes to nonlinear resonances
controlled by their geometry, i.e., the contour length around the end
of the tube and the length of the tube along its axis. It is quite
probable that the same mechanism applies also to microtubules in
biology. For the connection between ``quantum jumps" and nonlinear
resonances in classical phase space see Holthaus and Just \cite{hj}}

I would like to point out that the GRW-type localization
corresponds to a weak coupling limit of Hamiltonian systems with
coherent/squeezing interaction with the environment
\cite{ag}. Indeed Gaussian localization is specific to coherent and
squeezed states in the configuration representation. An immediate
scope is to generalize this type of localization to relativistic
quantum mechanics (RQM), and to quantum field theories (QFT).
In NRQM one is dealing with spatial probabilities, that is with
probabilities associated with a spatial domain ($\Delta X$) at a moment
of time T. To go to RQM, one must extend the spatial probability to a
spacetime domain as in \cite{y}.

\section{ Nonlinear wavefunction collapse ?}

The quantum wavefunction varies in time in a continuous way, following
the deterministic Schr\"{o}dinger evolution. When an observer wishes to
measure a physical quantity of a quantum system, the wavefunction
corresponding to that physical quantity is
exposed to an apparatus
specially designed to do this. The general effect of the apparatus,
usually macroscopic with respect to the physical system, is to induce a
discontinuous change
of the wave function from a superposition of states into just one
 state. This general effect is known in the quantum formalism as the
 collapse of the wavefunction. The open question is to find out the
 general mechanism of the collapse of the quantum wavefunction. In the
 literature one can find many interesting ideas on this problem.
As a quite acceptable interpretation of the collapsing
phenomenon we mention here the old ideas of Schr\"{o}dinger, who tried
to relate the modulus of the wavefunction to a materialistic and
realistic density of electronic matter, and {\em not} to probabilities.
For a recent discussion of this viewpoint the reader is referred to a paper
of Barut \cite{baru}. This model for the modulus of the wavefunction
can be elaborated  further by making use of progress due to Chew \cite{ch}.

In the following, we would like to comment on some phenomenological features
of physical collapses from other areas of physics in the hope to gain more
insights in the possible physical picture of quantum mechanical collapses of
admittedly fundamental origin. Our standpoint is that the present status of
the wave function reduction phenomenon is too formal, even though one may find
an abundant literature with interesting presentations of the topic \cite{coll}.
It is fair to say that we have no generally accepted physical mechanism of
the reduction process for the time being. In the literature, one can find
only extreme descriptions, claiming for a strong nonlinear process in which
gravity \cite{g} and/or quantum gravity \cite{qg} is thought
to play an important role. On the other hand, collapsing phenomena,
presumably displaying similar patterns can be encountered in several
other fields of physics, in the case when nonlinear effects are not balancing
any more the dispersive spreading of waves (solitons). Of course, in such
cases one is already outside the
restricted regime of linear dissipation implied by standard
quantum mechanics. Moreover, one can avoid thermodynamical arguments
against nonlinear variants of Schr\"{o}dinger equation \cite{p} by
making use of more general entropies \cite{m}.

A relevant example of nonlinear collapse is the Langmuir collapse
 in plasma physics. Langmuir collapses belong to the class of
 wave collapses, a well-defined topic in nonlinear physics \cite{non}.
The collapse of Langmuir wave packets in two or more dimensions
was first predicted by Zakharov \cite{z}, and it is observed in
the laboratory. It is a strong non-linear collapse occuring in
strong Langmuir turbulence, which consists of many locally
coherent wave packets interacting with a background of long-wavelength
incoherent turbulence \cite{rn}. Langmuir collapses are governed by
a non-linear Schr\"{o}dinger equation of the type
$ i\psi _{t} +1/2 \Delta \psi + |\psi |^{s} \psi =0  $
 which, as it is well-known, allows singularity
 formation in a finite time $t=t_{0}$, for $sd\geq 4 $, ($d$ is the
 dimension of space). The phenomenology of the Langmuir turbulence
 is extremely interesting. Wave packets are observed to
 ``nucleate" in existing density depressions. The nucleation of new
 wave-packets takes place by the trapping of energy from
 long-wavelength background turbulence into localized eigenstates
 of relaxing density wells. Since the collapse transfers energy to
 short scales, where there is strong damping, a process called
 ``burnout" occur in which energy is transfered to the electrons
 and the collapse is stopped. In this way the Langmuir field is
  dissipated, the density cavity relaxes and can serve as a nucleation
site for a new wave packet. Perhaps an equivalent physical picture as that
of the turbulent wave collapse might be made available with some modifications
for wavefunction collapse in a nonlinear scheme of quantum mechanics (e.g.,
a dust plasma model).

\section{Remarks on various other topics}

\subsection{Friction modifications of quantum mechanics}

Modifications of quantum mechanics may be thought of in terms of
friction terms for the more general situation of open quantum systems.
The problem of the ways of including various types of
friction in the quantum mechanical framework has been an active
field for decades. Many authors considered the dissipation in the form of
friction as a means to reconciliate quantum mechanics and general relativity,
and also as able to cast light on the transition between classical and
quantum physics. Even though the dissipation of energy seems to be
more appropriately described in terms of a density operator approach,
there has been always a steady activity towards understanding
friction at the level of wave functions \cite{dalc}.

In this area, the damped harmonic oscillator is considered to be
``the primary textbook example of the quantum theory of irreversible
processes", to quote Milburn and Walls \cite{mw}.

Some time ago, Ellis, Mavromatos and Nanopoulos \cite{emn} studied string
theory models from the frictional point of view. They gave reasons to believe
that the light particles in string theories obey an effective quantum
mechanics modified by the inclusion of a quantum-gravitational friction
term, induced by the couplings of the massive string states.
According to these authors the string frictional term has a formal
similarity to simple models of environmental quantum friction.

Finally, Beciu \cite{bc} sketched a proof showing that a friction term
for a cosmological fluid still retaining the symmetries of a perfect fluid
at the level of the stress tensor is equivalent to an inflaton field.

\subsection{Wave-particle dualities}

Historically speaking, the wave-particle dualities were established before
the advent of the quantum differential equations. We say dualities and not
duality because, not only for historical reasons, one must distinguish
the duality of photons from that of massive particles, say electrons.

The wave-particle duality of light is defined by the Einstein relation
$ E=h\nu=\frac{hc}{\lambda}$.
This duality of light was used by Einstein to explain the photoelectric
effect, by the Nobel-prize formula for the kinetic energy of the
emitted electrons $\frac{1}{2}mv^{2} =E-E_{0}$
where $E=h\nu$ is the quantized energy of the incident photons and
$E_{0}=e\phi$ is the threshold energy with $\phi$ the work function.

The duality of massive particles, on the other hand, was established by
de Broglie two decades after Einstein's duality. The wavelength and the
momentum of an electron (and of any other massive particle) is given by
$\lambda =\frac{h}{p}$.

It is worth noting the fact that the two dualities are related to
each other through the photoelectric effect,
$\frac{\hbar ^{2} k^{2}}{2m} =\frac{hc}{\lambda} -e\phi $.

Usually, the textbooks and the literature present the wave-particle
 dualities as a logical result of Young slit experiments.
As a rule, a more or less detailed discussion of the complementarity
 principle is accompaning the discussion of the Young experiment.
 Interesting ideas concerning the slit complementarity and duality
 have been put forth by Wootters and Zurek \cite{wz}, Bartell
 \cite{bar} and Bardou \cite{ba}. These authors made attempts to
 transcend the rather dogmatic presentation of this fundamental topic.
 Bartell introduced the idea of intermediate particle-wave behavior.
 Most probably, we need generalizations of the concepts of wave and
 particle, of their interactions, and a deep scrutiny of the effects
 of the type of experiment.

In the last couple of years, the investigation of particle-wave dualities
became a very active one, mainly because of the rapid progress of some
new technologies. Perhaps, one of the most interesting experiments
is that performed by Mizobuchi and Ohtak\'{e} \cite{mo}, which is
just a repetition of the old double prism experiment done by Bose
as long ago as 1897, however not with microwaves but, following a
suggestion of Ghose, Home, and Agarwal \cite{gha}, with single photon
states. An {\em and}-logic for the
wave-particle duality at the single-photon level has been claimed.

An open problem in detecting photons is the precise meaning of the
photon in the detection process. The point is that we are detecting
signals, and these signals depend on the experimental detection
schemes. The signals will give some pulses in the detectors.
Thus the  full detection process is governed by some electronic
relationships in the signal-pulse-detector system \cite{haj}.

Understanding better the manifestations
of wave-particle dualities for light can be highly relevant in photonics and
optical computing \cite{oco}.

At this point, let me quote from the recent paper entitled ``Anti-photon"
of W.E. Lamb, Jr. \cite{lamb}

\begin{quote}
... there is no such thing as a photon. Only a comedy of errors and
historical accidents led to its popularity among physicists and optical
scientists.

\end{quote}

Then, of course, the wave-particle duality for light will be loosing its
physical picture but will gain in mathematical rigor.

\subsection{The problem of the constancy of the Planck constant}

It was remarked by Barut \cite{b} that the free electromagnetic field
has no scale. There are only frequencies. Moreover, Planck originally
derived his formula from the properties of the oscillator on the
boundary of the black-body cavity, not from the quantization of the
field. The common practice of quantization of the fields came later.
 Therefore, we believe that even today careful experimental
checks of the constancy of the Planck constant should be made, and
in fact have been made in some laboratories \cite{f8}.
Barut showed that a formulation of quantum mechanics
without the fundamental constants $\hbar$, $m$ and $e$ is possible
\cite{b} \cite{w}. It looks like a pure wave theory in terms of
frequencies alone, and it might be
used more profitably in experiments where one measures frequency
differences. In this case, the energy becomes a secondary concept,
and different quantum systems are characterized by an intrinsic proper
frequency $\omega _{0}$. On the other hand, one can consider
quantum dynamics with two Planck constants, like did Di\'osi \cite{disi}.

As soon as we depart from the assumption of the constancy of the
Planck constant by merely considering a variable Planck parameter
(${\cal H}$), but nonetheless preserving the constancy of ${\cal H}/m$
 we may consider some kind of quantization at
large scales, planetary or even galactic ones. In fact there is a
quite vast literature on megaquantum effects. We draw attention to the fact
that such effects are related to the interpretation of ${\cal H}/m$ as
a pseudo-Planck constant which is associated to some gravitational
 systems (e.g., the Solar System \cite{bag}, quasars \cite{sar}).

A viewpoint to be recorded was put forward by Landsman \cite{ld}.
He claimed that only dimensionless combinations of $\hbar$ and a
parameter characteristic of the physical system under study are variable
in Nature. The references \cite{bag} \cite{sar} seem to confirm this idea.

We would like to point briefly on the possible effect of the spatial
scale of the measurement scheme on the numerical value of fundamental
constants. We shall use as an example the fine-structure constant
$\alpha =e^{2}/\hbar c$. At the present time, we know a very precise
macroscopic phenomenon,
namely the quantum Hall effect, from which the fine-structure constant can be
obtained from the quantized Hall resistance. (I consider quantized Hall
resistance a more precise experiment as compared to that involving the
proton gyromagnetic ratio, proton magnetic moment and Josephson
frequency-to-voltage ratio).
The numerical value obtained from the quantum Hall effect is
\cite{ct}: $\alpha ^{-1}=137.0359943(127)$. On the other hand,
the standard atomic measurement (coming from the anomalous magnetic
moment of the electron) gives $\alpha ^{-1} =137.0359884(79)$.
The two values differ only at the level of 0.1 ppm. The QED
corrections are confined to distances of the order of the Compton
wavelength of the electron, whereas the primary interaction in
the quantum Hall effect is between the electrons in the metal and those
circulating in the coils which produce the magnetic field. The
spatial scale in this case is of the order of a few cm. It would
be extremely interesting to relate the very small differences in
the numerical values of the fundamental constants to the spatial
scale of the phenomena used to measure them. Presumably, there might be
correlations between the last different digits of the numerical values
of the fundamental
constants and the spatial scale of the measuring device used to determine
that value. At least some self-similar correlations are to be expected.

\subsection{Quantum mechanics and cosmology}

The previous subsection already introduced us into the much more ambitious
program of describing the universe as a whole in quantum mechanical
terms. The difficult problem of interpretation is not so much with
respect to considering the Hilbert space of the Universe. It is related
to the fundamental fact that there can be no {\it a priori} division into
{\it observer} and {\it observed}. In other words, there is no
Feynman's ``rest of the Universe".

A generalization of the Copenhagen interpretation such as to be applied
to cosmology was first provided by Everett \cite{ev} in 1957. His
theory of ``many worlds" has been replaced at the present time by theories
of ``many histories" (time-ordered sequences of projection operators
\cite{gmh}), but the essential ideas remained those of 1957. As a matter of
fact, what Everett has done may be entailed in the process of probabilistic
 modeling, i.e., the organization of the space of wave function(s) as
 a probability space \cite{sg}.

 Everett showed how in
 his interpretation it is possible to consider the observer as part of
 the system (the universe) and how its fundamental activities-
 measuring, recording, and calculating probabilities- could be described
 by quantum mechanics. As incomplete points in Everett's interpretation,
 which has been much clarified subsequently, one should mention the
 origin of the classical domain we see all around us, and a more
 detailed explanation of the process of ``branching" that replaces the
 notion of measurement.

The main concept that has emerged in this area is that of decoherence
functionals, and the main debated topic is that of connecting this concept
to the probability interpretation.  Recently, Isham and collaborators
presented a
classification of the decoherence functionals based on a histories analog of
Gleason's theorem \cite{ish}. To be noticed are
the ``negotiations" on the border between quantum and classical in the
decoherence framework published in {\em Physics Today} of April 1993.

\subsection{Quantum jumps}

The interesting topic of quantum jumps \cite{qj}
has to do with
the rare but strong fluctuations that may show up in any stochastic
process, be it classical or quantum. The mathematical theory of large
deviation estimations has been already elaborated in considerable
extent \cite{des}.
All quantum mechanical equations have solutions to which
probability representations may be given \cite{kmc}.
The mathematical problem is to find out
probability measures of Poisson processes with jump trajectories,
which are similar to the Feynman-Kac transformation of probability
measures for processes with continuous trajectories. For relativistic
equations we have usually Poisson probability representations,
whereas for nonrelativistic equations diffusions in imaginary time
have been worked out, but also Poisson representations are possible.
One can establish the scale at which the transition from the covariant
hyperbolic Dirac dynamics to the non-covariant parabolic dynamics of the
Schr\"{o}dinger equation occurs \cite{sb}.

As a further argument that quantum jumps, i.e., ``discontinuities in time
of the wavefunction" in the terminology of Zeh \cite{z9}, are related to
rare fluctuations of stochastic nature, we remark that they are observed
even in single quantum systems \cite{e9}.

\subsection{Analogies to quantum mechanics}

Thinking by analogy is considered to be a clear indication of superior
reasoning and of human intelligence \cite{ana}.
 In physics there are a vast amount of analogies of much
help in the progress of many different branches of this science.
Many analogies are not complete and it is precisely this point to
induce into error all those residing too much on this beautiful
aspect of human thinking. One should keep in mind the danger of
extrapolating the analogies beyond their natural limits, which should
be carefully estimated, and also the risk of using them in the wrong
way.

Coming to quantum mechanics, we would like to recall
two quite attractive analogies. The first one is the electric
network discussed
by Cowan \cite{cow} long ago. The Cowan networks have the distribution of
the electric energy density in three dimensional space similar to that of
probability density waves corresponding to a spinless particle in
any potential field.

The second analogy has been recently discussed by J.L. Rosner \cite{ros}
who showed that the so-called Smith Chart method used for antenna
 impedance
matching corresponds in quantum mechanics to a simple conformal
transformation of the logarithmic derivative of the wavefunction.
The Smith Chart is a convenient graphical representation for analysing
transmission lines \cite{car}, and clearly may help understanding from
a different point of view the tunneling processes.

\subsection{Human brain and quantum computers/brains }

The flux of literature tells us that {\em quantum computers} are at good
moments of the {\em gate} phase and of exploratory discussions of various
physical setups from the quantum computational standpoint.
This exciting topic has been started
about two decades ago (though one can think of Szilard, von Neumann, and
Brillouin as well) and might turn into a really major general discipline.

Apparently the functioning of the human brain is not based
on quantum effects. The membrane voltages of the neurons do not imply the
Planck constant, and the important physical processes are essentially the
mesoscopic transport ones. A great advantage of the human brain is a quite
flexible microtubule architecture due to a remarkable phenomenon,
the so-called {\em dynamic instability} \cite{MK}. The origin of this
important phenomenon is debatable, and after having read the note of
Sumter and Noid (J. Chem. Phys. of April 22, 1995) I think that a nonlinear
resonance mechanism should be considered as a good proposal.
Many brain mysteries
are hidden in the microtubule assembly characterizing any individual
biological brain, and there is much unexplored physics.

The mesoscopic functioning of the human brain does not imply that an almost
quantum (e.g., nanoscopic) brain cannot be
fabricated. For example, Josephson junctions may be the component units
of such a brain since the relationship between the applied voltage and the
emitted frequency involves Planck's constant.

In his paper ``Is quantum mechanics useful ?" \cite{lan},
Professor Landauer remarked that {\em technologies differ in their explicit
utilization of quantum mechanical behaviour}.
The important technological task in considering quantum computers is to
print the bit on as small a material structure as physically possible in
order to
diminish the energy dissipation in the copying process, and to
substantially reduce the switching time from one bit to another. Actually,
the real
technological effort is evolving at the intricate nanometer scale, which
clearly will be essential for the general human progress. The emphasis
on the devices is this time both to understand what they measure and
mostly to estimate their computing capabilities.
As mentioned by
Feynman \cite{qco1} the present transistor systems
dissipate $10^{10}$ kT.
He considered
bits written ``ridiculously", as he said, on a single atom. At present
we know this is not ridiculous since we already are talking about atomic
transistors \cite{a}.

\section*{Acknowledgements}

I am grateful to Prof. Gian Carlo Ghirardi for encouraging me to participate
to the scientific activity on foundations of quantum theory in Trieste
along the summer and fall of 1992, when a first
substantial draft of this paper has been written.

This work was partially supported by the CONACyT Project 4868-E9406.


\end{document}